\documentstyle[buckow1,12pt]{article}
\input epsf.sty

\def\e{\epsilon}   
\def\g{\gamma}

\def\j{\psi}
\def\k{\kappa}     

\def\m{\mu}

\def\n{\nu}

\def\x{\xi}

\def\jb{\overline{\j}}

\newcommand{\del}{\partial}

\newcommand{\half}{\mbox{{\normalsize $\frac{1}{2}$}} }

\newcommand{\quart}{\mbox{{\small $\frac{1}{4}$}} }

\newcommand{\ra}{\rightarrow}

\newcommand{\lag}{\langle}
\newcommand{\rag}{\rangle}

\newcommand{\tk}{\widetilde{\kappa}}

\newcommand{\jn}{\j^{{\rm n}}}
\newcommand{\jbn}{\jb^{{\rm n}}}
\newcommand{\jc}{\j^{{\rm c}}}
\newcommand{\jbc}{\jb^{{\rm c}}}

\newcommand{\hmu}{\hat{\mu}}

\newcommand{\be}{\begin{equation}}
\newcommand{\ee}{\end{equation}}
\newcommand{\bea}{\begin{eqnarray}}
\newcommand{\eea}{\end{eqnarray}}
\newcommand{\eq}{\ref}
\newcommand{\beq}{\begin{equation}}
\newcommand{\eeq}{\end{equation}}

\newcommand{\lb}{\label}

\begin {document}
\def\titleline{
Gauge-Fixing Approach to Lattice Chiral 
\newtitleline
Gauge Theories 
}
\def\authors{
Wolfgang Bock \1ad , Maarten  F.L. Golterman \2ad , \\
Yigal Shamir \3ad                 
}
\def\addresses{
\1ad
Institute of Physics, Humboldt University, \\
Invalidenstr. 110, 10115 Berlin, Germany\\
\2ad
Department of Physics, Washington University  \\
St.~Louis,~MO~63130,~USA \\
\3ad
School of Physics and Astronomy, Tel-Aviv University \\
Ramat~Aviv,~69978~Israel
}
\def\abstracttext{
We review the status of our recent work on the gauge-fixing 
approach to lattice chiral gauge theories. New            
numerical results 
in the reduced version of a model with a U(1) gauge symmetry              
are presented which strongly indicate that the factorization 
of the correlation functions of the left-handed neutral 
and right-handed charged fermion fields, which we established 
before in perturbation theory, holds also nonperturbatively.
}
\makefront
\section{Introduction}
The nonperturbative formulation of chiral gauge theories 
on the lattice is a longstanding and, to date, still unsolved problem. 
The local chiral gauge invariance on the lattice is broken for 
non-zero values of the lattice spacing, even in models with an anomaly-free 
spectrum.
The failure of many proposals in the past was connected to the fact  
that the strongly fluctuating longitudinal gauge degrees of freedom alter the 
fermion spectrum leading to 
vector-like instead of chiral gauge theories. Already several years ago 
the Rome group proposed to use perturbation theory in the continuum 
as guideline and to transcribe the gauge-fixed continuum lagrangian of a chiral     
gauge theory to the lattice \cite{BoMa89}. 
The hope is that, for a smooth gauge-fixing condition (like e.g. the Lorentz gauge), 
preferably smooth gauge configurations are selected 
from a gauge orbit, whereas rough gauge field configurations 
are suppressed by a small Boltzmann weight.
The gauge-fixed model in the continuum is invariant under BRST symmetry. This 
symmetry is broken on the lattice, but the hope is that it can be restored 
by adding all relevant and marginal counterterms which are allowed
by the exact symmetries of the lattice theory.                 
Concrete lattice implementations of the nonlinear 
gauge $\sum_\m \{ \del_\m A_\m +  A_\m^2 \} =0$
and of the Lorentz gauge $\sum_\m \del_\m A_\m  =0$
were first given in refs.~\cite{Sh98} and 
\cite{GoSh96}.              

As a first step we have studied a model with U(1) gauge symmetry. 
The important advantage of the abelian case is that the ghost sector 
drops out from the path integral and no Fadeev--Popov term needs to be included 
in the action. As a second simplification we included only a 
gauge-boson mass counterterm which is the only dimension-two counterterm
and ignored all dimension-four counterterms.  
We notice already here that 
the lattice action can be formulated such 
that a fermion-mass counterterm (which is the only dimension-three 
counterterm) needs not to be added to the action. 
The coefficient of the gauge-boson mass counterterm has to be tuned such that the 
photons are massless in the continuum limit (CL). 
This value of the coefficient 
corresponds, for sufficiently small values of the gauge coupling, 
to a continuous phase transition between a ``ferromagnetic'' (FM) and a novel, 
so-called ferromagnetic ``directional'' (FMD) phase. The CL  
has to be taken from the FM side of the phase transition 
where the photon mass is larger than zero and the expectation 
value of the vector field vanishes. The FMD phase is characterized by 
a non-vanishing condensate of the vector field and a broken hypercubic 
rotation invariance.                                        

Motivated by previous investigations of lattice chiral gauge theories 
we first restricted the gauge fields to the trivial orbit, where 
only the dynamics of the longitudinal gauge degrees of freedom is taken into 
account \cite{Sh96r}.  
We shall refer to this model in the following as the ``reduced model."
In perturbation theory and 
by a high-statistics numerical simulation we could show in this reduced model 
that 
\begin{enumerate} 
\item the U(1)$_{\rm L, global} \otimes$U(1)$_{\rm R, global}$ is restored on the FM-FMD phase transition 
      which is a central prerequisite for the 
      construction of a chiral gauge theory on the lattice, and 
\item the fermion spectrum in the CL contains only the desired 
      left-handed charged fermion and a right-handed neutral 
      ``spectator'' fermion. 
\end{enumerate} 
The first statement applies 
also to the strongly coupled symmetric phase of the Smit-Swift model 
in which the unwanted species doublers 
were shown to decouple. The fermion spectrum in this phase  
however contains only a neutral Dirac fermion which 
decouples completely from the gauge fields when 
they are turned on back again. Later it was argued that a lattice chiral 
gauge theory can indeed not be defined    
within a symmetric phase or on its boundaries \cite{Sh93}. 

The outline of the rest of the paper is as follows: In Sect.~2. we 
introduce the fully gauged U(1) model and its reduced version. 
In Sect.~3 we review our previous results for the phase diagram 
and the fermion spectrum in the reduced model. 
Our new numerical results
which further substantiate the above statement about the 
fermion spectrum are presented in Sect.~4. 
In the last section, we  briefly summarize 
our results and give a brief outlook to future projects. 
\section{The Model}
The fully gauged U(1) lattice model 
is defined by the following action         
\bea
\!\!\! \!\! \!\!\! \!\!\!\!\!\!\!\!\!&& S_V=\frac{1}{g^2} \;  \sum_{x \m \n} 
\left\{ 1-\mbox{Re } U_{\m \n x} \right\}  
+ \frac{1}{2 \x g^2} \left\{ \sum_{x,y,z} \Box_{xy}(U) \Box_{yz}(U) -\sum_x B_x^2(V(U)) \right\} \nonumber \\
\!\!\! \!\! \!\!\! \!\!\!\!\!\!\!\!\!&&-\k \sum_{\m x} (U_{\m x} + U_{\m x}^{\dagger})
+\sum_{x,y} \left\{ \jb_x \gamma_\mu \left( [D_\m (U)]_{xy} P_L + [\del_\m]_{xy} 
P_R \right) \j_y  
-\frac{r}{2} \; \jb_x \Box_{xy} \j_{y} \right\}   \lb{SV} \\ 
\!\!\! \!\! \!\!\! \!\!\!\!\!\!\!\!\!&& B_x(V(U))=\quart \; \sum_\m ( V_{\m x-\hmu} + V_{\m x} )^2 \;, \;\;
V_{\m x} =  \mbox{Im}  \; U_{\m x}  \;. \lb{BX} 
\eea
The action in eq.~(\eq{SV}) 
includes the following terms (from the left to the right):    
the usual plaquette term ($\propto 1/g^2$), the Lorentz gauge-fixing term 
($\propto 1/2 \x g^2$), the 
gauge-boson mass counterterm ($\propto \k $), the ``naive'' kinetic 
term for the fermions and the Wilson term ($\propto r $) 
which we use to remove species doublers.
$U_{\m x}=\exp(i g A_{\m x})$ is the lattice 
link variable, $U_{\m \n x}$ the plaquette variable, 
$g$ is the gauge coupling, $\x$ is the gauge-fixing parameter, 
$r$ is the Wilson parameter 
and $P_{\rm L,R}=\half (1 \mp \g_5)$ 
are the left-and right-handed chiral projectors.    
$\partial_\m$ and $D_\m(U)$ designate the free and covariant antihermitian 
nearest-neighbor lattice derivatives, and $\Box$ and $\Box(U)$ 
the free and covariant nearest-neighbor lattice laplacians. 
The gauge-fixing action on the lattice was constructed such 
that it reduces in the classical CL to 
$\frac{1}{2 \xi} \int d^4 x 
(\del_\m A_\m)^2$, and has an absolute minimum at $U_{\m x}=1$, validating 
weak coupling perturbation theory \cite{Sh98,GoSh96}. We also notice that 
the fermionic part of (\eq{SV}) is invariant under 
the shift symmetry $\j_R \ra \j_R +\e_R$, $\jb_R \ra \jb_R +\overline{\e}_R$ 
\cite{GoPe89}. 
This symmetry implies that a fermion-mass counterterm 
does not need be added 
to the action. We ignore here all dimension four counterterms, which 
we believe to be less important \cite{GoSh96}. 
The lattice model is defined by the following 
path integral 
\be
Z=\int {\cal D} U {\cal D}\jb {\cal D}\j \; e^{-S_V(U;\j_L,\j_R)} 
=  \int {\cal D}\phi {\cal D} U {\cal D}\jb {\cal D}\j \; e^{-S_H(\phi;U;\j_L,\j_R)}                         \;.
\lb{PATH} 
\ee
In the second equation we have made the gauge degrees of freedom 
$\phi$ explicit. 
These gauge degrees of freedoms are nothing but group-valued scalar (Higgs)
fields. 
The action $S_H$ is obtained from $S_V$ by 
performing in eq.~(\eq{SV}) a 
gauge rotation $U_{\m x} \ra \phi^{\dagger}_x U_{\m x} \phi_{x+\m}$,
$\j_{{\rm L}x} \ra \phi_x^{\dagger} \j_{{\rm L}x}$. 
The scalar fields $\phi_x$ emerge in all those terms which are gauge 
non-invariant, i.e. in the gauge-fixing term, the gauge-boson mass 
counterterm and the Wilson term. $S_H$ is invariant 
under  a  U(1)$_{\rm L, local} \otimes$U(1)$_{\rm R, global}$ symmetry: 
$\j_{\rm L} \ra g_{{\rm L} x} \j_{{\rm L} x}$,
$\j_{\rm R} \ra g_{\rm R} \j_{{\rm R} x}$,
$U_{\m x}  \ra g_{{\rm L} x} U_{\m x} g_{{\rm L} x+\hmu}^{\dagger}$,
$\phi_{x}  \ra g_{{\rm L} x} \phi_x g_{\rm R}^{\dagger}$.

Next, we introduce the ``reduced" model, which is obtained by setting 
$U_{\m x}=1$ in $S_H$. 
The U(1)$_{\rm L, local} \otimes$U(1)$_{\rm R, global}$ symmetry turns into 
a U(1)$_{\rm L, global} \otimes$U(1)$_{\rm R, global}$ symmetry.           
The action of this reduced model reads           
\bea
 \!\!\! \!\!\!\!\!\! \!\!\!\!\!\!\!\!\! \!\!\!\!\!\! && 
S_H^{\rm red} =
 \tk \left\{ \sum_{x} \phi^{\dagger}_x (\Box^2 \phi)_x - B_x^2(V^r(\phi)) \right\}
-\k \sum_{x} \phi_x^{\dagger} (\Box \phi)_x
+\half \sum_{x \m } \left\{ \jb_x \gamma_\mu \j_{x+\m}  \right. \nonumber \\ 
 \!\!\! \!\!\!\!\!\! \!\!\!\!\!\!\!\!\! \!\!\!\!\!\! && 
-\jb_{x+\m} \g_\m \j_x  
\left. -r (( \jb_x (\phi_{x+\m}^{\dagger} P_L + \phi_x P_R) \j_{x+\m} + 
\mbox{h.c.}) -2 \jb_x 
(\phi_{x}^{\dagger} P_L + \phi_x P_R) \j_{x}) \right\}
\;, \lb{RED} 
\eea
where $V^r_{\m x}=\mbox{Im}(\phi_x^{\dagger} \phi_{x+\m})$.
The reason why this reduced model is of interest is that it 
should lead in the CL to a theory of free chiral fermions in the
correct representation of the gauge group. This is a necessary condition 
for the construction of chiral gauge theory with unbroken symmetry.
The failure of many previous proposals
of chiral gauge theories, like e.g. of the 
Smit-Swift model, was connected to the fact that the fermion spectrum is altered 
by the strongly fluctuating gauge degrees \cite{Sh96r}.    
In the following sections we shall reexamine this important question in  our 
model.                     
\section{Phase Diagram and Fermion Spectrum of the Reduced Model}
Let's first consider the phase diagram of the pure bosonic part of the action 
(\eq{RED}) which only includes the gauge-fixing term and the gauge-boson mass 
counterterm. The $(\k,\tk)$-phase diagram of this higher derivative scalar field 
theory contains at large $\tk$ 
a FM phase at $\k > \k_{{\rm c}}$, where 
$\lag V^r_{\m x} \rag = 0$ and $\lag \phi \rag > 0$,
and a FMD phase at $\k < \k_{{\rm c}}$ with 
$\lag V^r_{\m x} \rag \neq 0$ \cite{Sh98}. Both phases are separated by a continuous phase 
transition at $\k_{{\rm c}}(\tk)$ \cite{BoGoSh97a}. 
To one-loop order we find $\k_{{\rm c}} = 0.02993 + O(1/\tk)$ \cite{BoGoSh97a}.
As explained in Sect. 1, 
in the fully gauged model $\k$ has to be tuned from the FM side 
towards this phase transition in order to obtain massless photons.  
In the reduced model we have computed 
the order parameter $\lag \phi \rag$ in the FM phase, both 
in perturbation theory in $1/\tk$ and also numerically, and find that it vanishes 
in the limit $\k \searrow \k_{\rm FM-FMD}$ \cite{BoGoSh97a,BoGoSh98}. 
This phenomenon is associated with the 
$1/(p^2)^2$ propagator for the $\phi$-field fluctuations.
The vanishing of $\lag \phi \rag $ implies that the 
U(1)$_{\rm L, global} \otimes$U(1)$_{\rm R, global}$ 
symmetry which is broken to its diagonal subgroup in the FM phase,
is restored on the FM-FMD phase transition line, an essential 
prerequisite for the construction of a chiral gauge
theory with unbroken gauge symmetry. 

We now introduce the four 
fermion operators $\jn_{\rm R}=\j_{\rm R}$, $\jn_{\rm L}=\phi^{\dagger} \j_{\rm
L}$,
$\jc_{\rm L}=\j_{\rm L}$ and $\jc_{\rm R}=\phi \j_{\rm R}$.
The fields with the superscripts c (charged)
and n (neutral) transform nontrivially under the U(1)$_{\rm L, global}$ and
U(1)$_{\rm R, global}$ subgroups respectively.
We have calculated the neutral and charged fermion
propagators both to one-loop order in perturbation theory in $1/\tk$ 
\cite{BoGoSh98}, 
and also numerically \cite{BoGoSh97b}. 
We could show that 1.) the unwanted species doublers
decouple in all cases and 2.) 
only the 
$\jn_{\rm R}$- and $\jc_{\rm L}$-propagators exhibit in the CL 
isolated poles at $p=(0,0,0,0)$, 
which correspond to the desired massless fermion states. We  
found that non-analytic terms occur in  
the $\jn_{\rm L}$- and $\jc_{\rm R}$-propagators
and that there are no poles in these channels 
in the CL.  If the $\jn_{\rm R}$- and $\jc_{\rm L}$ fermions 
are the only free fermions that exist in the CL of              
the reduced model, we would expect that the $\jn_{\rm L}$-and 
$\jc_{\rm R}$-correlation functions in coordinate space 
{\em factorize} for sufficiently large separations $|x-y|$ in the following manner 
\be
\lag \jn_{{\rm L},x} \jbn_{{\rm L},y} \rag  \sim 
    \lag \jc_{{\rm L},x} \jbc_{{\rm L},y} \rag 
   \lag \phi_{x}^{\dagger} \phi_{y} \rag  \;,\;\;\;\;\;\;\;\;\;
 \lag \jc_{{\rm R},x} \jbc_{{\rm R},y} \rag  \sim 
     \lag \jn_{{\rm R},x} \jbn_{{\rm R},y} \rag \lag \phi_{x} \phi_{y}^{\dagger} \rag  \;.
\lb{FAC} 
\ee
We were able to show in one-loop perturbation in $1/\tk$ 
that the two relations in eq.~(\eq{FAC}) hold both in the FM phase and also 
in the CL, i.e. for $\k \searrow \k_{\rm FM-FMD}$ \cite{BoGoSh98}.
It is important to confirm these relations also nonperturbatively.
\section{Numerical Results}
To this end we have performed a 
quenched simulation at the point $(\tk,\k)=(0.2,0.3)$ in the FM phase.
We set $r=1$ and  determined the five correlation functions 
$\lag \jn_{{\rm L},x} \jbn_{{\rm L},y} \rag$,
$\lag \jn_{{\rm R},x} \jbn_{{\rm R},y} \rag$,
$\lag \jc_{{\rm L},x} \jbc_{{\rm L},y} \rag$,
$\lag \jc_{{\rm R},x} \jbc_{{\rm R},y} \rag$ and 
$\lag \phi_{x}^{\dagger} \phi_{y} \rag$ 
on an cylindrical lattice of size $L^3 T$ with $L=6$, $T=24$.
For the fermion fields
we used antiperiodic (periodic) boundary conditions in the
temporal (spatial) directions, whereas for the scalar field
we used periodic boundary conditions in all directions.

To compute the four fermionic correlation functions numerically we 
have to employ point sources and sinks 
which implies that a very high statistics is required 
to obtain a satisfactory signal to noise ratio. 
We set 
\be 
y(x,t)=x_1 \hat{1} + x_2 \hat{2} +x_3 \hat{3} 
+ \mbox{mod}(x_4+t,T) \hat{4} \;,  \lb{SINK}
\ee
with $t=1,\ldots,T-1$.  For a given scalar field configuration 
we have randomly picked a source point 
on each time slice and averaged over 
the resulting $T$ correlation functions. 
For the computation of the fermionic correlation functions we used  
in total 1300 scalar configurations which were generated
with a 5-hit Metropolis algorithm and, in order to reduce autocorrelation 
effects, were separated by 2000 successive Metropolis sweeps.     

In the case of the bosonic correlation function we 
summed, for a given configuration, over all lattice sites $x$, 
$ \lag (1/(L^3T) \sum_x \phi_{x}^{\dagger} \phi_{y(x,t)} \rag$, where $y(x,t)$ is 
given again by  eq.~(\eq{SINK}), 
and measured the bosonic correlation function 
on each of the $1300 \times 2000$ configurations.               
%
\begin{figure}
\vspace*{-0.9cm}  
\begin{tabular}{cc}
\hspace{-1.0cm} \epsfxsize=9.00cm
\epsfbox{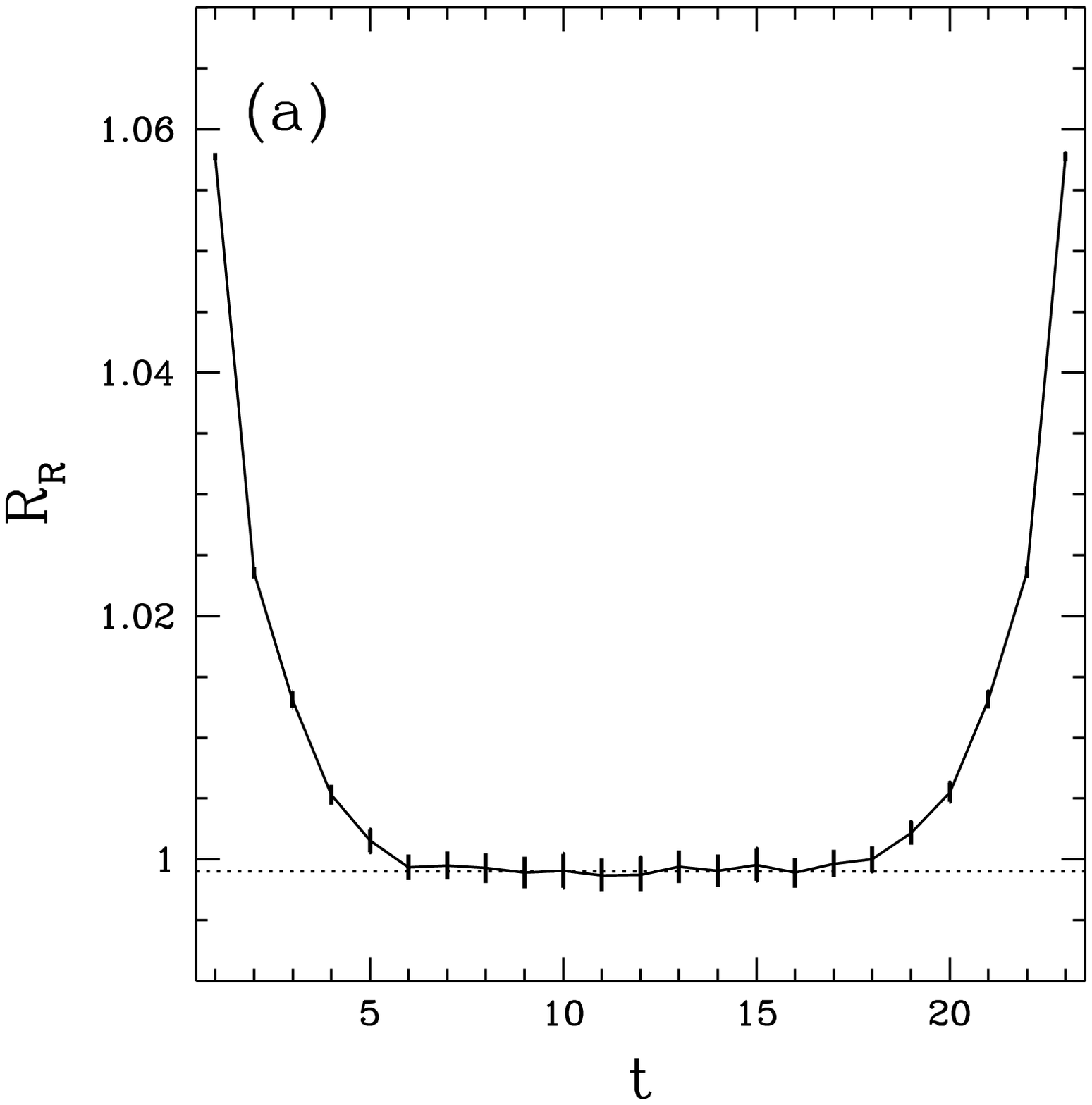} &
\hspace{-2.3cm}
\epsfxsize=9.00cm
\epsfbox{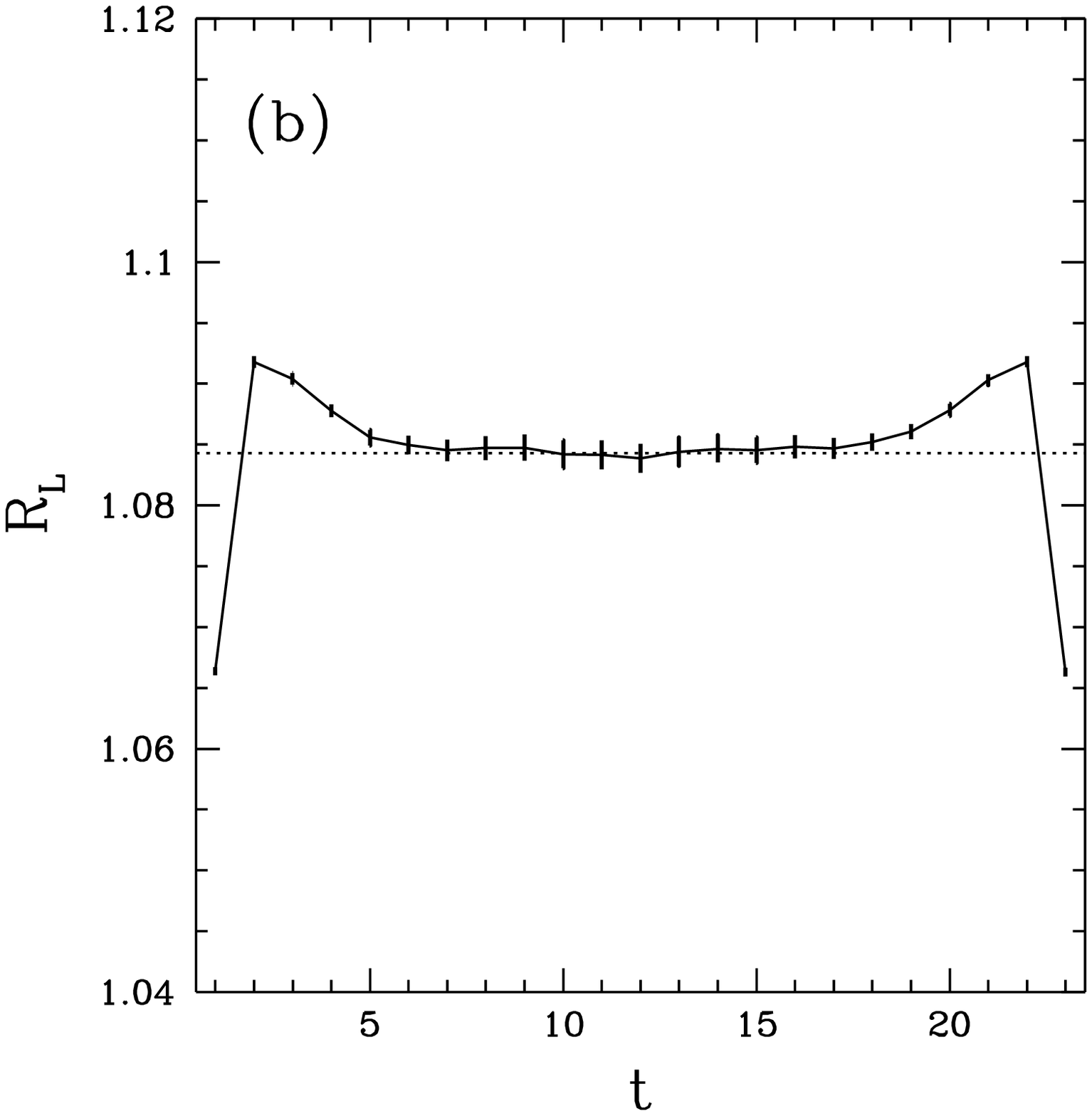}
\end{tabular}
\vspace*{-1.4cm} \\
\caption{The ratios $R_R$ (a) and $R_L$ (b) as function 
of $t$ at $(\tk,\k)=(0.2,0.3)$ ($r=1$). The lattice size is $6^324$. 
The data points are connected by solid lines and 
the horizontal dotted lines are to guide the eyes. 
}                   
\label{FIG1}
\end{figure}
%

To check if relation (\eq{FAC}) holds also nonperturbatively 
we have computed the two ratios 
\be
R_L=\frac{  \lag \jn_{{\rm L},x} \jbn_{{\rm L},y} \rag }
   { \lag \jc_{{\rm L},x} \jbc_{{\rm L},y} \rag 
   \lag \phi_{x}^{\dagger} \phi_{y} \rag } \;,\;\;\;\;\;\;\;\;\;
R_R=\frac{  \lag \jc_{{\rm R},x} \jbc_{{\rm R},y} \rag }
   {  \lag \jn_{{\rm R},x} \jbn_{{\rm R},y} \rag \lag \phi_{x}^{\dagger} \phi_{y} \rag } \;,
\lb{RATIO} 
\ee
which should approach a constant at sufficiently large separations 
$t$. The two ratios are displayed in fig.~\ref{FIG1} as a function 
of $t$. The two graphs clearly show 
that both ratios start to flatten off at $t \approx 5$ 
and are, within errors, indeed constant at larger separations. 
The fact that $R_R=1$ (for larger t) is consistent with
shift symmetry. 
\section{Conclusion}
The quenched results presented in the last section suggest  
that the factorization  
of the $\j^{\rm n}_L$-and  $\j^{\rm c}_R$-correlation functions 
(cf. eq.~(\eq{FAC}))
which we established 
before in 1-loop perturbation theory (cf.~ref.~\cite{BoGoSh98}) 
remain valid also nonperturbatively. 

As future direction of research we plan to 
study the U(1) model with full dynamical gauge fields. This
requires the fermion representation to be anomaly free.     
We furthermore want             
to extend the gauge-fixing approach to the nonabelian
gauge theories. This is a non-trivial issue, 
because it is not known whether the
BRST formulation of gauge theories can be defined consistently
beyond perturbation theory.  \\

\noindent {\em Acknowledgements}: 
WB is supported by the Deutsche 
Forschungsgemeinschaft under grant Wo 389/3-2, MG by 
the US Department of Energy as an Outstanding Junior Investigator,
and YS by the US-Israel Binational Science
Foundation, and the Israel Academy of Science.
%

\end{document}